\documentclass{article}

\usepackage[utf8]{inputenc} 
\usepackage{microtype}
\usepackage{graphicx}
\usepackage{subfigure}
\usepackage{booktabs} 

\usepackage{hyperref}

\usepackage[accepted]{icml2020}

\icmltitlerunning{Capsule Neural Network for Tuberculosis detection}

\begin{document}

\twocolumn[
\icmltitle{Using Capsule Neural Network to predict Tuberculosis \\ in lens-free microscopic images}



\icmlsetsymbol{equal}{*}

\begin{icmlauthorlist}
\icmlauthor{Dennis Núñez-Fernández}{one}
\icmlauthor{Lamberto Ballan}{two}
\icmlauthor{Gabriel Jiménez-Avalos}{one}
\icmlauthor{Jorge Coronel}{one}
\icmlauthor{Mirko Zimic}{one}
\end{icmlauthorlist}

\icmlaffiliation{one}{Laboratorio de Bioinformática y Biología Molecular, Universidad Peruana Cayetano Heredia, Peru}
\icmlaffiliation{two}{Visual Intelligence and Machine Perception Group
University of Padova, Italy}

\icmlcorrespondingauthor{Mirko Zimic}{mirko.zimic@upch.pe}

\icmlkeywords{Machine Learning, ICML}

\vskip 0.3in
]



\printAffiliationsAndNotice{}  

\begin{abstract}
Tuberculosis, caused by a bacteria called Mycobacterium tuberculosis, is one of the most serious public health problems worldwide. This work seeks to facilitate and automate the prediction of tuberculosis by the MODS method and using lens-free microscopy, which is easy to use by untrained personnel. We employ the CapsNet architecture in our collected dataset and show that it has a better accuracy than traditional CNN architectures.
\end{abstract}

\section{Introduction}

Tuberculosis (TB), caused by the Mycobacterium tuberculosis bacteria, is one of the most serious public health problems in Peru and worldwide. Around a quarter of the world's population, are infected with TB and 1.2 million people died from TB in 2018 \cite{1_}.

The MODS method, (Microscopic Observation Drug Susceptibility Assay) \cite{2_} developed in Peru, allows the quick recognition of morphological patterns of mycobacteria in a liquid medium (directly from a sputum sample). It is a cost-effective option, since TB can be detected in just 7 to 21 days, which is a fast and a low-cost alternative test. This method is included in the list of rapid tests authorized by the National Tuberculosis Prevention and Control Strategy. In addition, complementing this method with lens-less microscopy is very important due it is easy to use and to calibrate by no trained personnel.

In recent years Convolutional Neural Networks (CNNs), a class of deep neural networks, have become the state-of-the-art for object recognition in computer vision \cite{3_}, and have potential in object detection \cite{5_} and segmentation \cite{3_}.

For the analysis of medical images via deep learning, image classification is often essential. Manual evaluation by experienced clinicians, is important; however, it is laborious and time-consuming, and may be subjective. Another approach for image classification, a fully convolutional neural network known as CapsNet \cite{30_}, has recently shown promising results \cite{6_}. The CapsNet has been applied for some medical and microscopy images, including classification of 2D HeLa cells in fluorescence microscopy, classification of Apoptosis in phase contrast microscopy, and classification of breast tissue \cite{10_,8_,11_}. 

The purpose of this work is to evaluate the feasibility of automatic TB detection on lens-free images. We applied the CapsNet architecture for image classification.

\section{Methodology}

For prediction of TB on grayscale images, the proposed methodology makes use of image patching followed by the CapsNet classifier in order to have an efficient classifier using few training images. We have the following steps: First the full image (3840x2700 pixels) is divided into sub-images (256x256 pixels), then each sub-image is passed forward the CapsNet classifier and provides the positive/negative label for each sub-image. Then, the histogram of classes for every full image is constructed. Finally, given a full image, the logistic regressor makes the prediction of TB.

\begin{figure}[H]
\vskip 0.2in
\begin{center}
\centerline{\includegraphics[width=0.9\columnwidth]{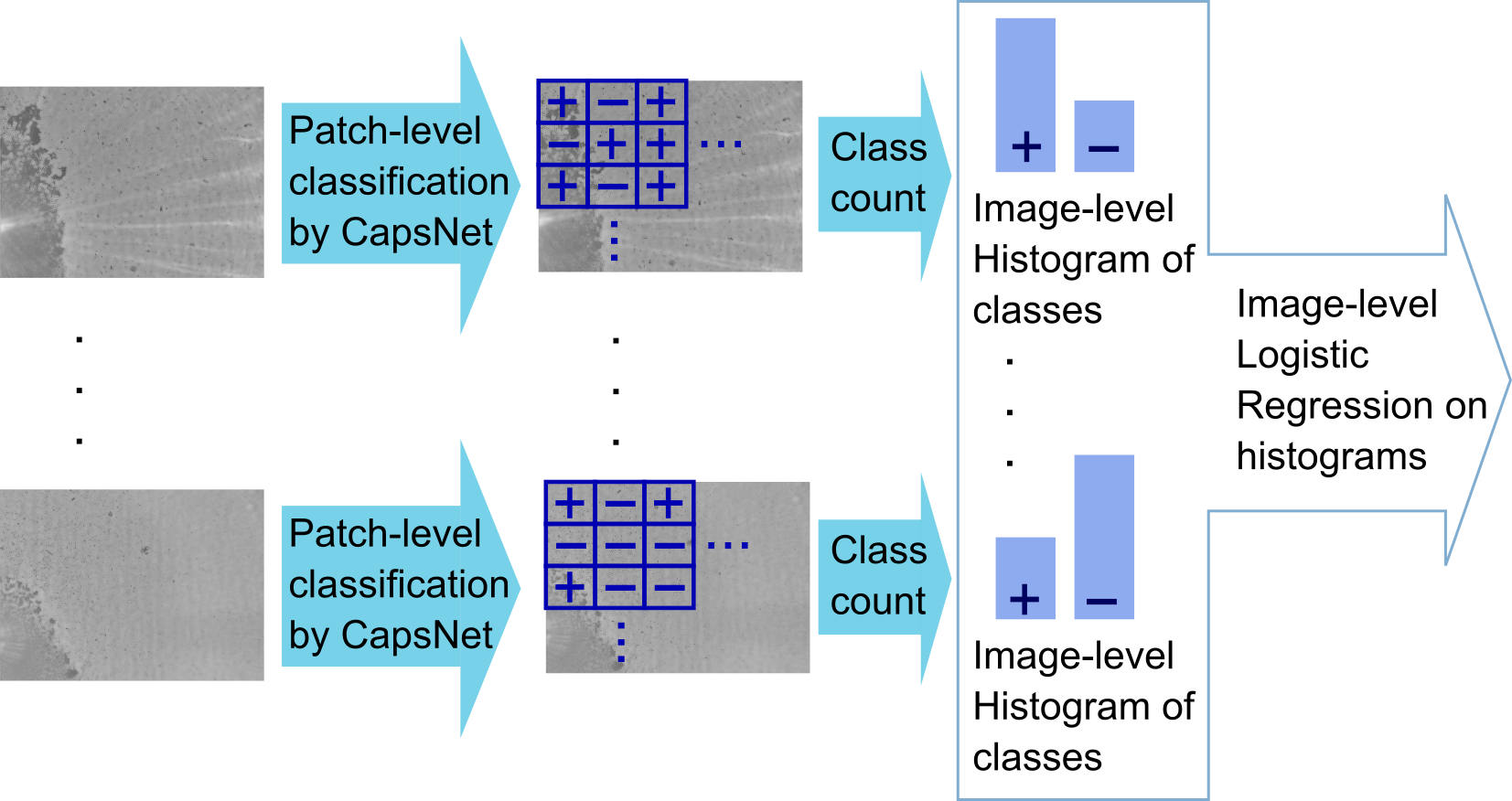}}
\caption{Overview of the methodology.}
\label{overview}
\end{center}
\vskip -0.2in
\end{figure}

\begin{figure}[H]
\vskip 0.2in
\begin{center}
\centerline{\includegraphics[width=0.90\columnwidth]{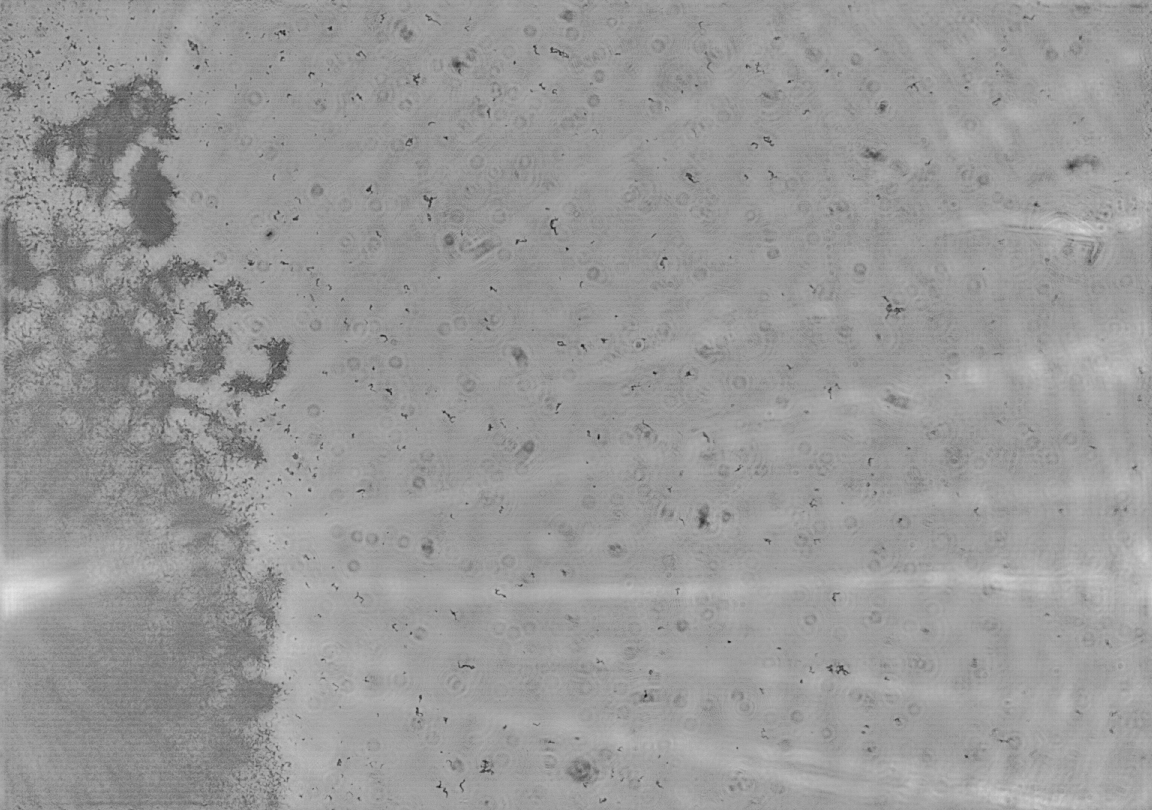}}
\caption{Sample of a single full image.}
\label{sample}
\end{center}
\vskip -0.2in
\end{figure}

\begin{figure}[H]
\vskip 0.2in
\begin{center}
\centerline{\includegraphics[width=0.65\columnwidth]{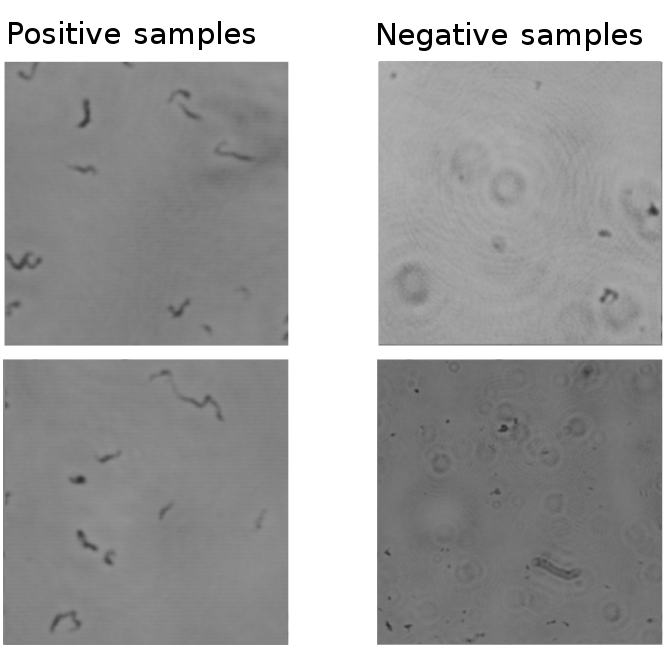}}
\caption{Samples of sub-images of the collected dataset.}
\label{samples}
\end{center}
\vskip -0.2in
\end{figure}

The images were collected by the iPRASENSE Cytonote lens-free microscope. These images were reconstructed from the holograms that were obtained by the microscope via the iPRASENSE software. As shown in Figure~\ref{sample}, the collected images have a dimension of 3840x2700 pixels and are presented in grayscale format. After data collection and manual annotation, we obtained 10 full annotated images. To augment this dataset and due the images are very large compared to the TB cords, we performed the division of the images into sub-images. So, each image was divided into sub-images of 256x256 pixels. Since some TB cords could be near the margin of each sub-image, we considered an overlapping of 20 pixels between adjacent sub-images. With the above mentioned considerations and taken the same number of images for both classes (areas without Tb cords in the full images are larger than the areas with TB cords), we obtained a total of 500 positive and 500 negative sub-images. These positive and negative images were used to train/test the different models, see Figure~\ref{samples}.

The annotation of the collected images was as follows: First a medical expert, which has extensive experience in TB, visually inspected each sub-image. If the medical expert found at least one TB cord the sub-image was annotated as positive, otherwise annotated as negative.

\begin{figure}[H]
\vskip 0.2in
\begin{center}
\centerline{\includegraphics[width=0.9\columnwidth]{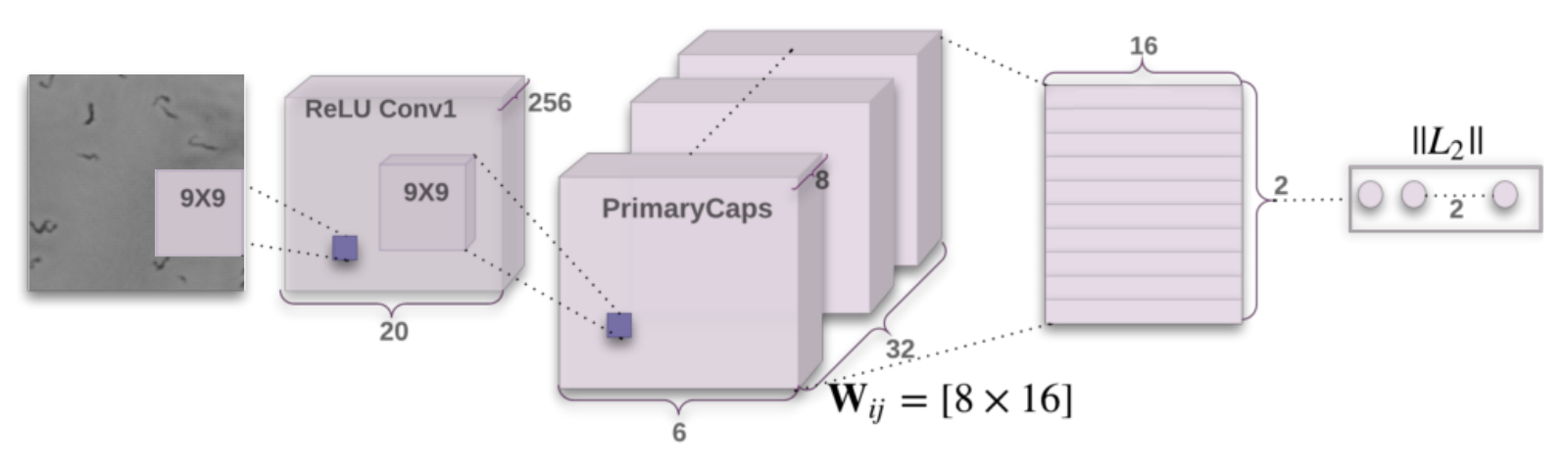}}
\caption{Architecture of the propesed CapsNet.}
\label{capsnet}
\end{center}
\vskip -0.2in
\end{figure}

The network is based on the CapsNet convolutional network, see Figure~\ref{capsnet}. There, the first two layers of the convolution kernel we used for the training model are 9*9. The selection of the convolution kernel is related to whether the low-level features can be accurately obtained. If the size of the convolution kernel is too small, the selected features will be enhanced accordingly, but if there is noise in the image, the noise will be enhanced, thus overwhelming the useful details in the image. 

In addition, the Capsule Network performance is relatively strong. On small datasets, CapsNet is significantly better than CNN. For very few samples, the capsule network still has good accuracy and convergence \cite{30_}.

\section{Results and Discussion}

The CapsNet was trained over 450 positive and 450 negative sub-images, and tested on 50 positive and 50 negative sub-images. In order to compare the CapsNet performance in the sub-images, the same classification task was done on different CNN architectures: LeNet \cite{31_}, AlexNet \cite{3_} and VGG-19 \cite{32_}. As shown in Table~\ref{cnns-table}, the CapsNet architecture gives a better accuracy. So, for all shuffle testing on the dataset, we obtain an accuracy of 84.7\%. We must consider that the main challenge is the dataset, which is noisy and has a low degree of resolution. The main reason why CapsNet gives better results is the fact that the structures of TB cords are floating over the sputum sample and obtaining different 3D shapes. This is where the Capsnet performs an essential task, as it is invariant to 3D position changes. In addition, the overall performance of the system based on 20 full images (10 TB positive and 10 TB negative) provides a sensitivity of 80\% using the CapsNet architecture for sub-image classification.

\begin{table}[H]
\caption{Classification for different CNN architectures for the same classification task.}
\label{cnns-table}
\vskip 0.15in
\begin{center}
\begin{small}
\begin{sc}
\begin{tabular}{lcccr}
\toprule
\textbf{Approach} & \textbf{Accuracy} \\
\midrule
LeNet & 67.2$\pm$ 0.6 \\
AlexNet & 73.4$\pm$ 0.8 \\
VGG-19 & 79.7$\pm$ 1.4 \\
\textbf{CapsNet} & \textbf{84.3$\pm$ 1.2} \\
\bottomrule
\end{tabular}
\end{sc}
\end{small}
\end{center}
\vskip -0.1in
\end{table}

\section{Conclusions}

This work shows that the CapsNet architecture provides a classification accuracy of 84.3\%, which is better than other CNN architectures such as AlexNet of VGG-19, for TB prediction based on sub-image division. Also, this paper demonstrates that our system is capable for prediction of TB with a sensitivity of 80\%, this by using images captured from a lens-free microscopy and using the CapsNet architecture. The results of the test seem promising despite the fact that the images are noisy and low quality. As described in this work, our project contributes to the development of a fast, cost-effective, and globally useful screening tool for tuberculosis detection. The development of such systems helps to improve medical solutions in general, but especially in rural areas where medical resources are limited.

\section*{Acknowledgements}

This project has been financed by CONCYTEC, by its executing entity FONDECYT, with the objective of promoting the exchange of knowledge between foreign academic institutions and Peru.


\bibliography{references}
\bibliographystyle{icml2020}

\end{document}